\documentclass[journal,article,submit,pdftex,moreauthors]{Definitions/mdpi}

  \newcommand{\vbr}{\vec{r}}
  
  \newcommand{\vbk}{\vec{k}}
  \newcommand{\vba}{\vec{A}}
  \newcommand{\vbf}{\vec{F}}
  \newcommand{\vbd}{\vec{d}}

\newcommand{\infi}{\ensuremath{\mathrm{if}}}
\newcommand{\vbpi}{\vec{\Pi}}
\newcommand{\Eref}[1]{Equation (\ref{#1})}
\newcommand{\eref}[1]{Eq.~(\ref{#1})}

\firstpage{1} 
\pubvolume{1}
\issuenum{1}
\articlenumber{0}
\pubyear{2026}
\copyrightyear{2026}
\datereceived{ } 
\daterevised{ } 
\dateaccepted{ } 
\datepublished{ } 

\Title{Strong-field Photoionization: Analysis of Overlapping Above-Threshold Ionization and Laser-Assisted Photoemission Structures }

\Author{Candelaria Migliaro $^{1}$, Juan~Martín Randazzo $^{1,2}$\orcidB{} and Renata Della~Picca $^{1,2}$\orcidA{}*}

\AuthorNames{Candelaria Migliaro, Juan~Martín Randazzo and Renata Della~Picca }

\address{%
$^{1}$ \quad Instituto Balseiro (UNCuyo), 8400 Bariloche, Argentina\\
$^{2}$ \quad Centro Atómico Bariloche (CNEA) and CONICET, 8400 Bariloche, Argentina}

\corres{Correspondence: renata@cab.cnea.gov.ar (R.D.P.)} 

\abstract{
We present a theoretical description of atomic strong-field photoionization. Specifically, we consider an atom driven by a combination of two electromagnetic fields: a high-frequency field assisted by an intense, low-frequency laser. We investigate the photoelectron spectrum (PES) as the sum of two contributions: direct ionization due to the laser field and the photoionization term associated with the high-frequency field.
We identify  the contributions of above-threshold ionization (ATI) and laser-assisted photoemission (LAPE) structures in the total spectra, even when they overlap. As a particular case, we investigate the situation where an ATI-peak coincides with a sideband. 
Our theoretical scheme for the hydrogen initially in the $1s$ quantum state and based on strong field approximation, is general enough to be applied to other atomic species and field configurations.
}


\begin{document}


\section{Introduction}

The interaction of strong and short laser pulses with atoms and molecules has received renewed attention, mainly due to advances in laser technology that have enabled new experimental investigations of atomic and molecular processes on ultrashort time scales and under extreme intensity conditions \cite{Joachain2012, Vrakking2024}.
When an atom or molecule is exposed to an intense infrared (IR) laser field, ionization can occur through above-threshold ionization (ATI), a process in which the system absorbs more photons than the minimum required for ionization \cite{Milosevic2006, Arbo2010, Joachain2012}. On the other hand, when an extreme ultraviolet (XUV) pulse overlaps in space and time with an IR laser field in the presence of matter, laser-assisted photoemission (LAPE) processes take place. Depending on the duration of the XUV pulse, two distinct scenarios arise: the sideband and the streaking regimes. Each of these contributes specific structures to the electron emission probability \cite{Gramajo2018, DellaPicca2020}.
It is important to note that, in the presence of LAPE, direct ionization processes induced by the IR field also occur, among which ATI is a prominent example. ATI and LAPE are fundamentally different processes and, under typical experimental conditions, give rise to well-separated structures in the photoelectron spectrum. In the usual scenario, ATI structures extend from threshold up to twice the ponderomotive energy, while the XUV contribution is approximately centered around the XUV photon energy. By appropriately choosing the XUV frequency, both spectral domains can be kept apart.

However, depending on the laser parameters, these structures may overlap. In this work, we explore such overlapping regimes. Several questions arise: Under which conditions can each structure be studied separately? Is it valid to treat them as independent? When does their superposition become relevant? And must this superposition be treated coherently?
To address these questions, we employ the strong-field approximation (SFA), a powerful analytical tool, and focus on the hydrogen atom H($1s$) as a tested system.

The paper is organized as follows. 
In Sec. \ref{theory}, we briefly resume the SFA theory of ATI and LAPE processes and analyze the properties of the photoelectron spectra (PES) for a linearly polarized probe IR field and a pump XUV pulse.
In~\ref{overlap}, we analyze the main features and behavior of the PES as the field parameters are varied,
identifying the domains where each structure (either ATI or LAPE) dominates, as well as their overlapping region. 
Concluding remarks are presented in Sec.~\ref{Conclusions}.
Atomic units are used throughout the paper, except where otherwise stated.

\section{Theory and Results} \label{theory}

Let us consider the ionization of an atomic target  by the combination of two EM fields: a high-frequency field (typically in the XUV regime, noted by $X$) assisted by another intense, low-frequency laser ($L$). 
The Hamiltonian of the system in the single active-electron approximation is $H=H_0+H_\textrm{int}(t)$, where $H_0$ is the time independent atomic Hamiltonian and the $H_\textrm{int}(t)$ describes the interaction of the atom with both time-dependent $X$ and $L$ EM fields in the dipole approximation and length gauge: $H_\textrm{int}(t)= \vbr\cdot[\vbf_X(t) + \vbf_L(t)] $.

The electron initially bound in an atomic state $\phi_{i}$ is emitted to a final continuum state 
with final momentum $\vbk$, energy $E=k^2/2$ and solid angle $\Omega$. 
Within the time-dependent distorted wave theory, the energy photoelectron spectra (PES) can be calculated as $d P/d E = \int d \Omega \  \sqrt{2 E} \ (d P/ d\vbk)$ where the differential emission probability is
\begin{equation}
\frac{d P}{d \vbk } =  |T_{\infi}|^2 , 
\quad \textrm{ and } \quad 
T_{\infi}= -i\int_{-\infty}^{+\infty} d t \,\langle\phi_{f}^{-}(\vbr,t)|H_\textrm{int}(\vbr,t)|\phi_{i}(\vbr,t)\rangle
\label{PES-TIF}
\end{equation}
is the $T$-matrix element corresponding to the transition $\phi_{i}\rightarrow\phi_{f}$ in the prior form \cite{Macri2003, DellaPicca2013}. 
The initial atomic state is $\phi_{i}(\vbr,t)=\varphi_{i}(\vbr)\,e^{i I_{p} t}$, with
ionization potential $I_{p}$, and $\phi_{f}^{-}(\vbr,t)$ is the distorted final state.
Within the Strong Field Approximation (SFA) the Coulomb distortion in the final channel is neglected as well as the
the influence of the laser field in the initial ground state. 
Hence, we can approximate the distorted final state with a Volkov function 
\cite{Joachain2012, symphonySFA}, \textit{i.e.}, $\phi_{f}^{-}(\vbr,t)= \chi_{f}^{V}(\vbr,t)$, where
\begin{equation}
 \chi_{f}^{V}(\vbr,t) = 
 (2\pi)^{-3/2} \exp(i \, \vbpi \cdot \vbr )
\times \exp \Big(  \frac{i}{2} \int_{t}^{\infty}  \vbpi^2 d t^\prime  \Big),
\end{equation}
$\vbpi(\vbk,t)= \vbk+\vba(t) $
 and the vector potential due to the total external field is defined as 
$\vba(t) = -\int^t_{-\infty} d t'[\vbf_{X}(t')+\vbf_{L}(t')]$.
Since the interaction Hamiltonian has two terms, the transition matrix can also be splitted into two contributions and written as:
\begin{equation}
T_{\infi} = T_X + T_L \qquad \textrm{ where } \qquad
T_{X(L)}  = -i\int_{-\infty}^{+\infty} \vbf_{X(L)}(t) \cdot \vbd[\vbpi(\vbk,t)]  \,  e^{i S(\vbk,t)} \, d t  \label{eqTif}
\end{equation}
for the $X$ or $L$ pulse respectively.
Here the generalized action $S$ and dipole moment $\vbd$ are defined as
\begin{equation}
S(\vbk,t) = I_p t +  \int_{\infty}^t \frac{\big[\vbpi(\vbk,t^\prime) \big]^2}{2}  dt^\prime
\qquad \textrm{ and } \qquad
\vbd(\vbpi)  =  \langle \frac{e^{i \vbpi \cdot \vbr} }{(2\pi)^{3/2}} | \vbr| \varphi(\vbr)\rangle.
\end{equation}
%
The total potential vector is the sum of two linearly polarized pulses of duration $\tau_i$, polarization $\hat{\varepsilon}_i$ and carrier frequency $\omega_i$ where $i=X$ or $L$ respectively; then $\vba = \vba_X +\ \vba_L$ where each term is modeled as
\begin{equation}
\vba_i  = \frac{F_{i0}}{\omega_i} \, f_i(t) \,
\sin\big[\omega_i(t-t_{0i})  \big]\, \hat{\varepsilon}_i .
\end{equation}
The proposed envelope function is  $f_i(t) =\cos^2\big[\pi(t-t_{0i})/ \tau_i  \big] $  only during the temporal interval $(t_{0i}-\tau_i/2, t_{0i}+\tau_i/2)$ and vanishes outside. 
Then, the electric field peak amplitude $F_{i0}$ is reached at the middle of the pulse, \textit{i.e.} at time $t_{0i}$. 
Aditionally, we consider each pulse duration as a integer number of its oscillation cycle: $\tau_i=  N_i ~ 2 \pi/ \omega_i $.

In the following, we first separately consider each term of the transition matrix \eref{eqTif} and its relationship with ATI and LAPE processes, to finally address the total spectrum as sum of both contributions simultaneously. 
In order to present specific results we perform our calculations for the case of hydrogen atom initially in the $1s$ state and for EM fields with polarization along $\hat{z}$.  
Our theoretical scheme can be applied, however, to
other atomic species and field configurations as well.

\subsection{ATI review}

By setting $F_{X0}=0$ in the previous section we recover the strong field ionization process by only the field $L$. Then, we introduce the transition matrix for ATI as: 
$T^\textrm{ATI} = T_L \big|_{F_{X0}=0}$.
It is well known that ATI takes place when the multiphotonic regime is achieved; then the atom can absorbs more photons that the minimum needed to ionize. 
The PES presents the well known ATI-peaks at energy values given by:
\begin{equation}
E_n = n \omega_L - I_p - U_p \label{eqATI}
\end{equation}
with $U_p= (F_{L0}/2\omega_L)^2$. 
There are several works that have studied the characteristics of ATI-peaks (height, width, shape, position, etc.) as a function of the laser profile, see for example \cite{Duchateau2002, Milosevic2006, Arbo2010, Joachain2012,  DellaPicca2013} for more details.

As an example, we present the PES for ATI process in Figure~\ref{fig1} in orange solid line for a particular case of ionization of H(1s). We observe the presence of 3 ATI-peaks corresponding to the absorption of 
2, 3 and 4 photons of frequency $\omega_L$.

\subsection{LAPE review}

When the high frequency field is also present the PES presents specific structures corresponding to the absorption of the high energy photon. 
In the sideband regime, it has been shown that the intercycle interference due to the coherent electronic emission at each optical cycle of the laser, gives rise to the sideband peaks in the PES at energy values 
\begin{equation}
E_m = \omega_X + m \omega_L - I_p - U_p \label{eqLAPE},
\end{equation}
corresponding to the absorption (positive $m$) or emission
(negative $m$) of $m$ laser photons, following the absorption of
one $X$ photon (see for example \cite{Gramajo2016, Gramajo2018, DellaPicca2020}).
What is more, this intercycle structure is modulated by the intracycle factor restricting the spectrum to the energy domain established by the semiclassical model between  limits
 $E_\textrm{low,up} = (v_0 \mp F_{L0}/\omega_L)^2/2$ (where $v_0^2/2=\omega_X-I_p $ corresponds to the mean energy of the photoelectrons ionized by the $X$ pulse in absence of the laser).

To analytically demonstrate these features, the authors of \cite{Gramajo2016, Gramajo2018, DellaPicca2020}  have 
worked out the transition matrix $T_X$ assuming that,
as the frequency of the $X$ pulse is much higher than the $L$ one,
 and considering the strength of the first one is much smaller than the laser one, 
the $X$ contribution
to the total vector potential can be neglected ($\vba_X\simeq 0$). 
Furthermore, within the rotating wave approximation which accounts for the absorption of only
one $X$ photon and neglects the contribution of its emission, the $X$ electric field becomes $\vbf_X(t)\simeq F_{X0}(t) e^{-i\omega_X t} \ \hat{\varepsilon}_X$.
Then, they have finally approximated $T_X$ by
\begin{equation}
T^\textrm{LAPE} = T_X \big|_{\vba=\vba_L}=-i\int_{-\infty}^{+\infty} F_{X0}(t)\hat{\varepsilon}_X  \cdot \vbd[\vbk + \vba_L(t)]  \,  e^{i S_{LX}(\vbk,t)} \, d t ,
\label{eqTLAPE}
\end{equation}
where 
$S_{LX}(\vbk,t)= (I_p -\omega_X ) t +   \int_{\infty}^t dt^\prime \big[\vbk+ \vba_L(t^\prime) \big]^2/ 2  $ and $F_{X0}(t)$ is the time dependent amplitude of the $X$ electric field.

\begin{figure}[H]
\includegraphics[width=0.7 \linewidth]{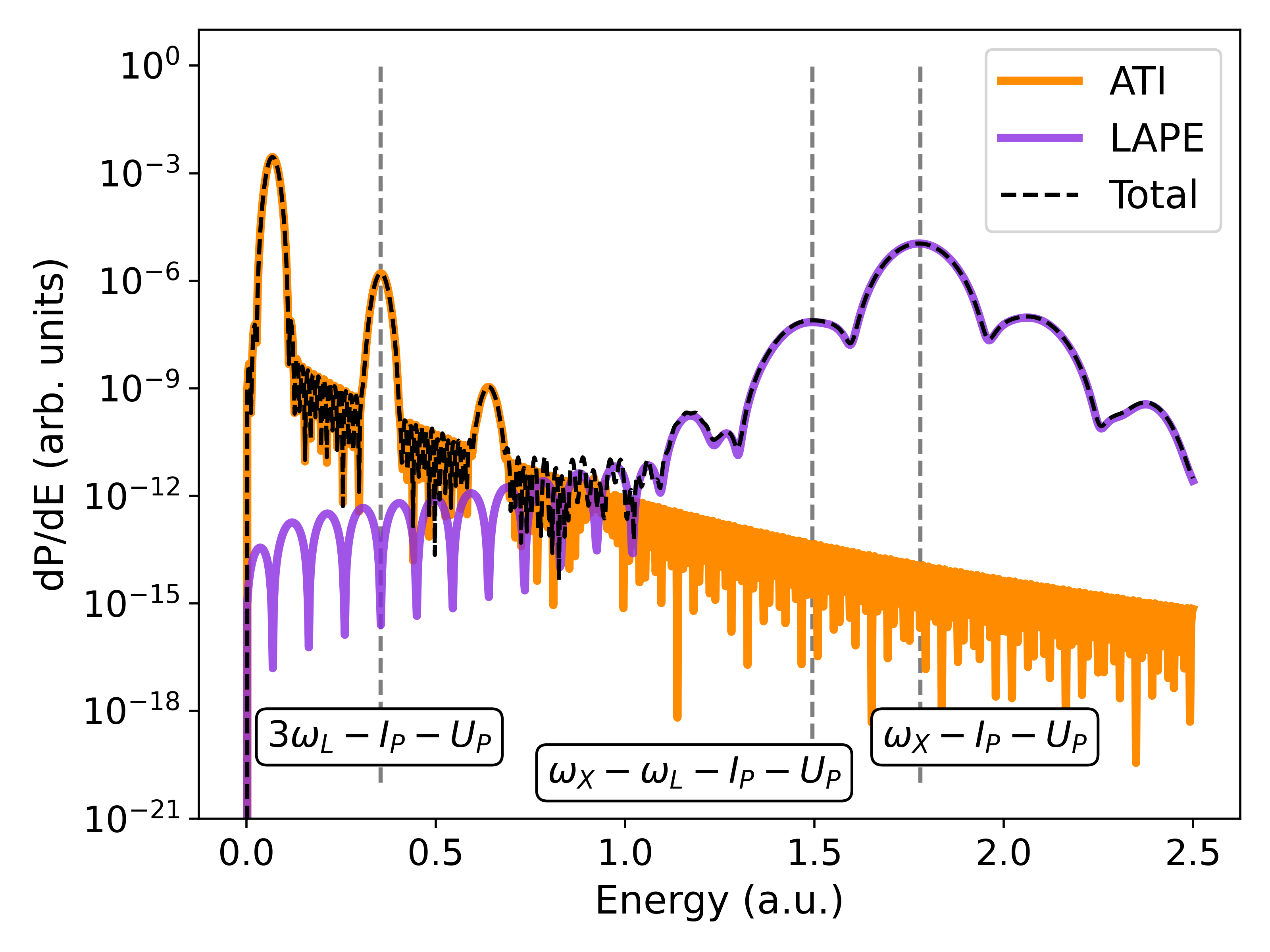}
\caption{PES for H(1s) as a function of the photoelectron energy considering only $T^{ATI}$ (orange line), only  $T^{LAPE}$ (purple line) and the coherent sum \eref{coherent} [idem \eref{incoherent}] in dashed line. 
The field parameters are  $\omega_L = 0.285$ a.u., $F_{L0} = 0.01$ a.u., $N_L$ = 20, $\omega_{X} = 8\omega_L$, $F_{X0} = 0.001$ a.u., $N_X = 24$ and $t_{0X} = t_{0L}$.
Vertical dashed lines indicate the energy position corresponding to the boxed value. 
\label{fig1}}
\end{figure}

In Figure~\ref{fig1} we show the LAPE spectra in solid purple line for the pulse parameters specified in its caption. We observe that 
the sideband peaks corresponding to $m=0,\pm 1, \pm 2$ can be clearly distinguished in the PES. 
We can also observe  that the energy domain of the LAPE process is well separated from the domain of ionization by the laser alone (orange line). 
In other words, the contribution of direct laser ionization is negligible in the energy
domain where the absorption of one $X$ photon takes place. 
In the following section we will vary the parameters of both pulses to study the superposition of structures in PES.

\subsection{Overlapping ATI and LAPE processes} \label{overlap}

 As we have written in \eref{PES-TIF} and \eref{eqTif},  due to the combination of both $X$ and $L$ pulses, the ionization probability corresponds to the coherent sum of $T_X$ and $T_L$. 
However, as we already discussed in the previous sections, with the appropriate choice of the $X$ and $L$ pulse parameters, each term can be approximated by $T^\textrm{LAPE}$ and $T^\textrm{ATI}$ respectively. Even more, if the spectra do not overlap in the energy domain, the sum can be made incoherently: 
\begin{eqnarray}
d P / d \vbk &=& \big| T_X + T_L   \big|^2 
\qquad \textrm{coherent sum} \label{coherent} \\
 &\simeq& \big| T^\textrm{LAPE}   \big|^2 + \big| T^\textrm{ATI}   \big|^2 \qquad \textrm{incoherent sum} \label{incoherent}
\end{eqnarray}


Provided the assumptions of the approximations made in \eref{incoherent} are satisfied, both \Eref{coherent} and \eref{incoherent} will yield equivalent results.
This is shown, for example, in Figure~\ref{fig1}, where the dashed line indicates that the total ionization probabilities (sum of the ATI and LAPE spectra, represented by orange and purple lines, respectively) produce exactly the same curve using both \eref{coherent} and \eref{incoherent}, both plotted with dashed lines.

\begin{figure}[H]
\begin{adjustwidth}{-\extralength}{0cm}
\centering
\subfloat[\centering]{\includegraphics[width=8.0cm]{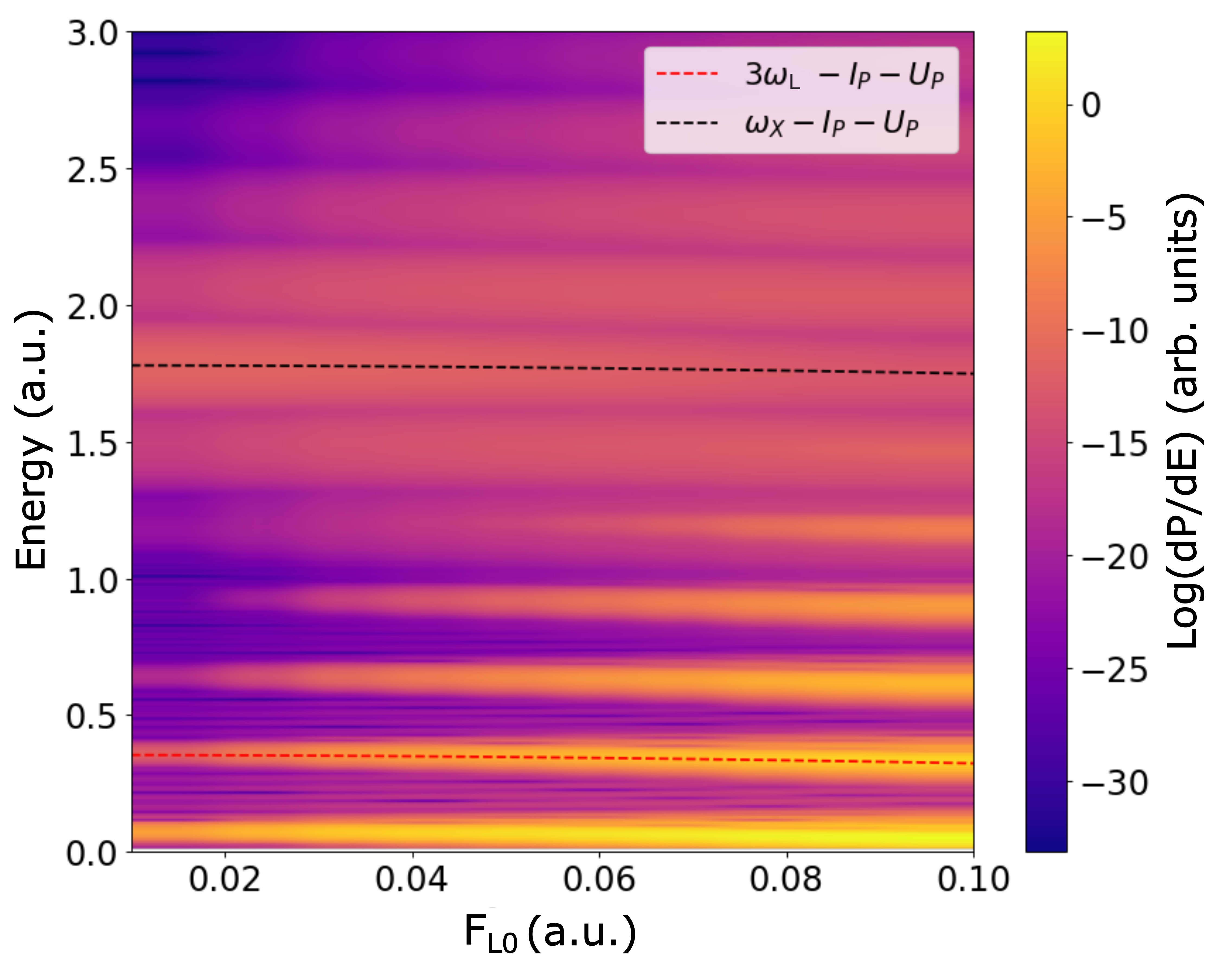}}
\subfloat[\centering]{\includegraphics[width=8.0cm]{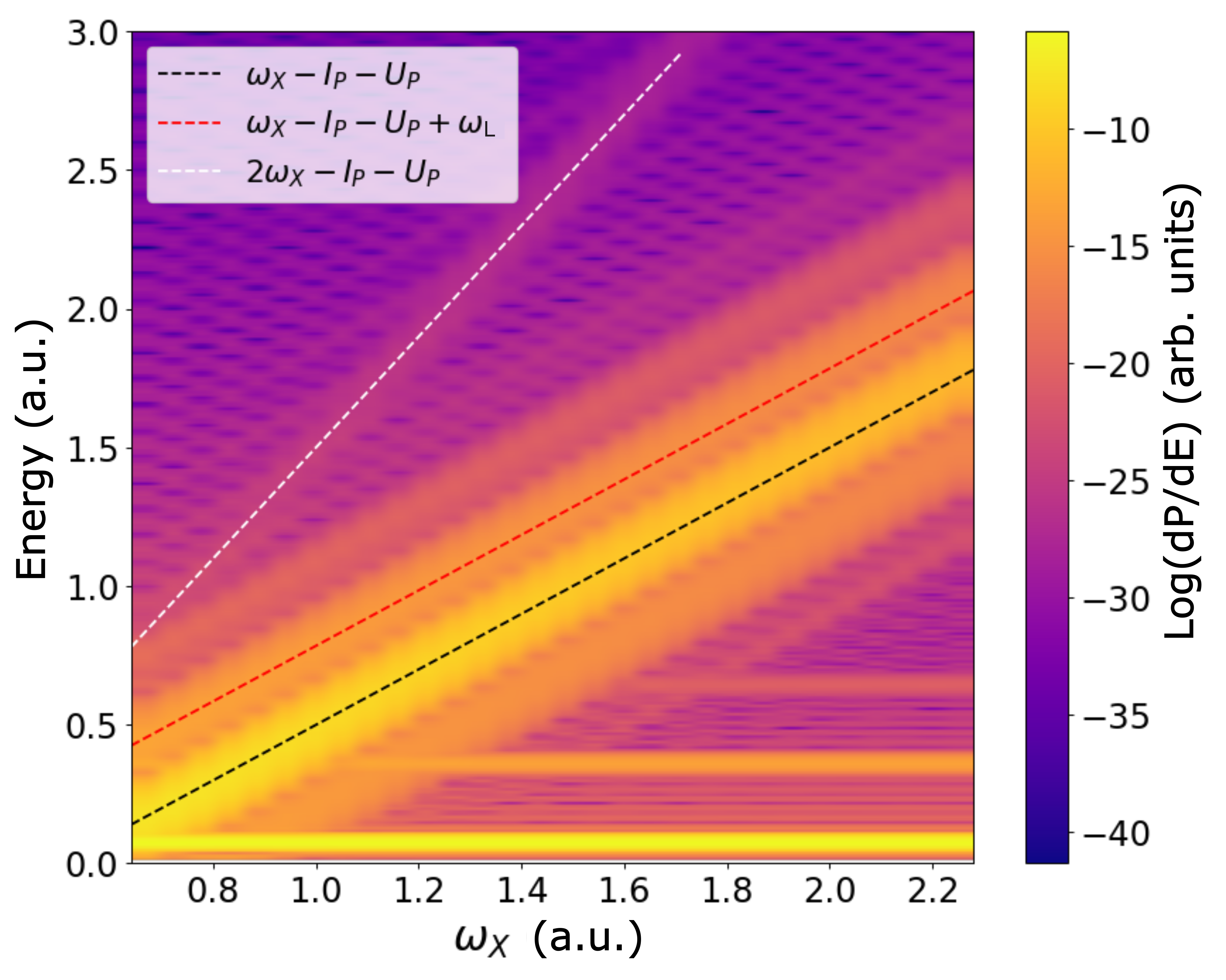}}
\end{adjustwidth}
\caption{H($1s$) PES in color scale for angle integrated \eref{coherent}:
(\textbf{a}) as a function of the laser peak amplitude $F_{L0}$ and
(\textbf{b}) as a function of the frequency $\omega_{X}$.
Other field parameters are idem Figure~ \ref{fig1}.
\label{fig2}}
\end{figure}

However, a reasonable question would be to inquire whether \eref{incoherent} remains valid in other parameter regimes. In order to explore it beyond its range of applicability, we present in Figure~\ref{fig2} the PES as a function of the amplitude $F_{L0}$ and frequency $\omega_X$.
The amplitude of the $L$ field is related to the width of the ATI and LAPE structures. Therefore, increasing the amplitude $F_{L0}$ also increases their domain. At a certain point, they may overlap.
This is precisely what is observed in Figure~\ref{fig2}(a): 
as $F_{L0}$ increases, more photons become available, giving rise to additional ATI-peaks and sidebands.
Meanwile for low amplitudes, ATI-peaks are spectrally separated from the $X$ photon absorption peak and its sidebands.
On the other hand, it can also be seen that the position of the peaks does not remain at the same energy; instead, they shift quadratically toward lower energies due to the ponderomotive shift $-U_p$ [see \eref{eqATI}].

Figure~\ref{fig2}(b) shows the PES  as a function of the frequency of the $X$ field. Since the position of the sidebands depends on $\omega_X$ [see \eref{eqLAPE}],  decreasing it allows the spectra to overlap. This fact can be seen in the Figure:  the structures that follow diagonal lines correspond to the absorption peak of the $X$ photon accompanied by its sidebands. 
In contrast, at low energies  the ATI-peaks remain constant, appearing as horizontal lines. 
A third line (white dashed line) with a steeper slope (2 a.u.) can also be observed. This maximum corresponds to the absorption of two $X$ photons. 

From Figure~\ref{fig2} we observe that, even when overlap occurs,  ATI and LAPE structures remain unchanged. 
In this sense, the total spectrum resembles the incoherent sum.
This seems to be the general rule: \eref{incoherent} satisfactorily reproduces \eref{coherent} in most of the investigated range; except obviously for the two $X$ photon zone.

\begin{figure}[H]
\begin{adjustwidth}{-\extralength}{0cm}
\centering
\subfloat[\centering]{\includegraphics[width=8.0cm]{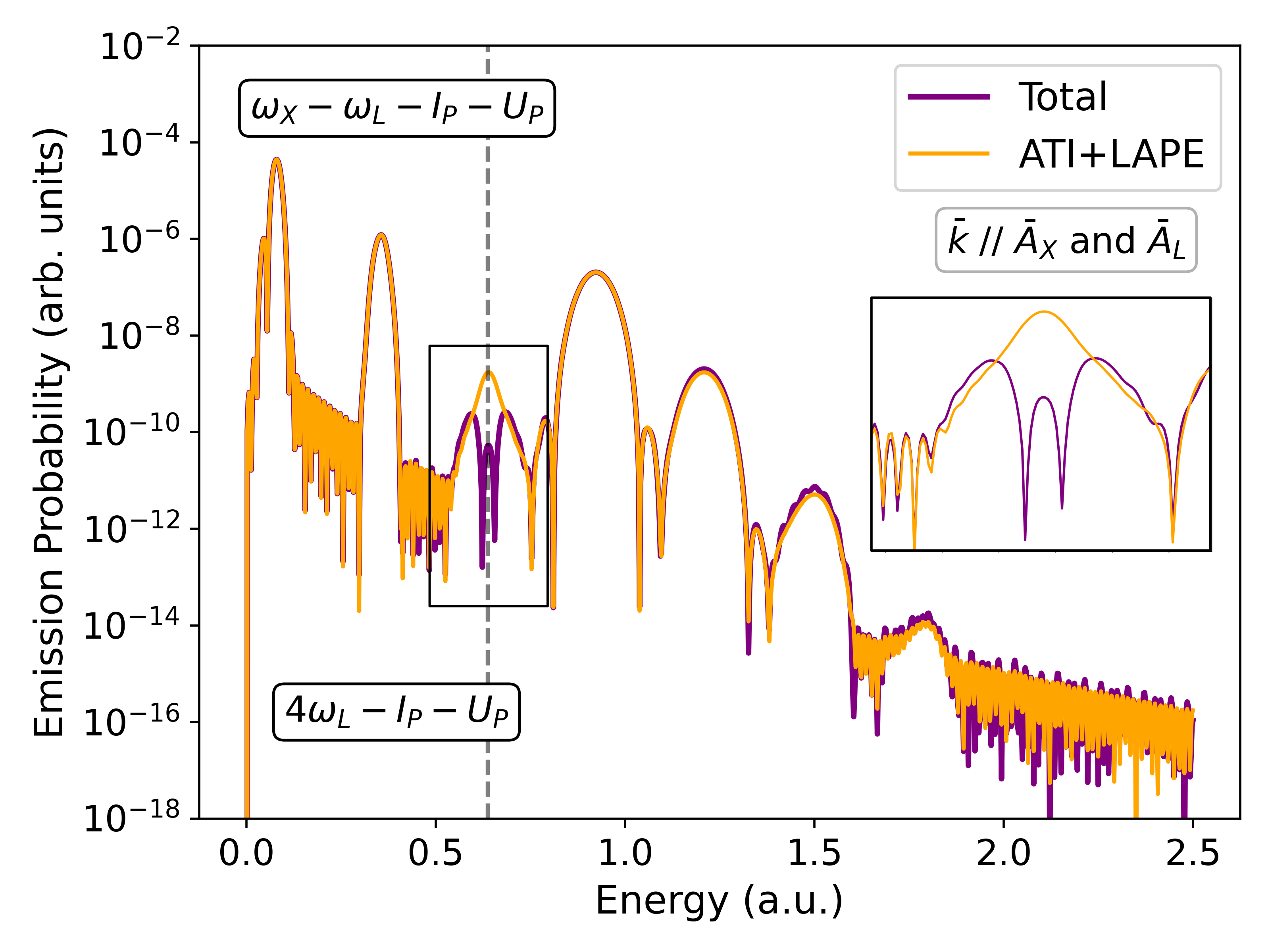}}
\subfloat[\centering]{\includegraphics[width=8.0cm]{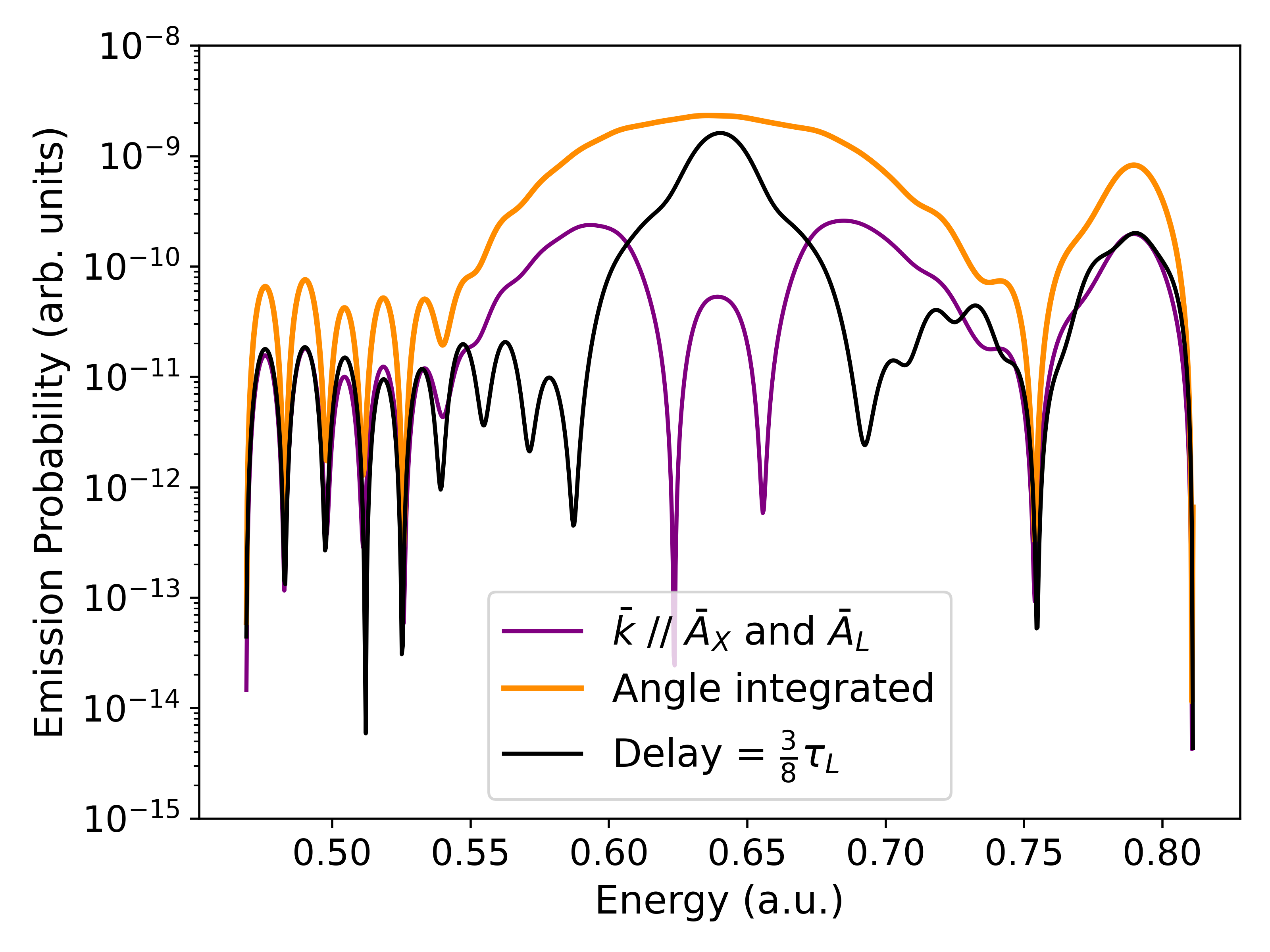}}
\end{adjustwidth}
\caption{Differential emission probability for H($1s$) corresponding to forward emission:  $\vbk \parallel \hat{z}$. 
Laser pulse idem to Figure~\ref{fig1}. 
For the $X$ field, the parameters are $\omega_{X} = 5\omega_L$ a.u., $F_{X0} = F_{L0}/150$, $N_X = 25$ and $t_{0X} = t_{0L}$.
(\textbf{a}) \eref{coherent} in purple line and \eref{incoherent} in orange line. 
(\textbf{b}) Zoom of panel (a): \eref{coherent} in purple line idem panel (a), coherent PES \eref{coherent} integrated in emission angles in orange line and in black line coherent \eref{coherent} with delay $t_{0X} - t_{0L} = \frac{3}{8}\tau_{L}$.
\label{fig3}}
\end{figure}

In what follows, we look for particular conditions under which a noticeable difference between the PES from \eref{coherent} and \eref{incoherent} arises. 
To this end, we aim for a scenario where a sideband peak coincides in the energy position of an ATI-peak, and both have comparable heights.
In Figure~\ref{fig3}(a), we show the spectra for an emission direction parallel to both polarization vectors. After carefully choosing specific parameters, we finally observed a case where the coherent and incoherent sums differ, even if only in a very localized region of the spectrum. 
In the coherently summed spectrum (purple line), one can observe the suppression of the peak at energy 0.64 a.u., which corresponds both to the ATI-peak with $n=4$ [see \eref{eqATI}] and to the $m=-1$ sideband [see \eref{eqLAPE}]. 
In contrast, \eref{incoherent} yields a simple sum of both contributions.
This could be interpreted as quantum interference between the different pathways leading to the same final energy: absorbing 4 photons of the laser, or absorbing one $X$ photon and emitting one $L$ photon.
Figure~\ref{fig3}(b) shows that upon angle integration, this destructive interference is erased. Moreover, a slight change in the delay between the two pulses 
also makes the interference disappear.
This behavior closely resembles that observed in a RABBIT protocol, where sidebands exhibit a suppression pattern depending on the delay between pulses \cite{Lopez2024, Ocello2025}. This finding requires deeper analysis, which will be explored in a future work.

\section{Conclusions} \label{Conclusions}

We have studied strong-field photoionization within the framework of the strong-field approximation (SFA), considering the electromagnetic field as the sum of two pulses: a high-frequency contribution (typically in the XUV regime) assisted by an intense, low-frequency laser (usually in the infrared domain). From the analytic properties derived from the SFA transition matrix element, we have identified the contributions of the above threshold ionization (ATI) and laser-assisted photoemision (LAPE) transition matrices and their corresponding structures in the total spectra.

After exploring a wide range of parameters, we conclude that, in general terms, the total spectrum corresponds to a simple superposition of both ATI and LAPE structures. However, upon closer inspection, in some very specific circumstances, quantum interference between different electron emission pathways can be observed, leading to destructive interference. In particular, we have analyzed the situation where an ATI-peak coincides with a sideband, resulting in the suppression of the electron emission probability. Slight variations in the laser parameters cause this destructive interference to disappear. The detailed study of this interference phenomenon will be the subject of future research.


\vspace{6pt}

\acknowledgments{The authors acknowledge support by National University of Cuyo (Grant No: 80020240100520UN, 2025 -2027).}


\isPreprints{}{
\begin{adjustwidth}{-\extralength}{0cm}
} 

\reftitle{References}

\end{adjustwidth}

\begin{thebibliography}{999}
%
\bibitem{Joachain2012}
Joachain, C.; Kylstra, N.; Potvliege, R.
\textit{Atoms in Intense Laser Fields}, 1rd ed.; Publisher: Cambridge University Press, UK, 2012. 
%
\bibitem{Vrakking2024}
Vrakking, M. 
Faster than a speeding bullet - the 2023 Physics Nobel Prize. 
{\em J. Phys. B: At. Mol. Opt. Phys.} {\bf 2024}, {\em 57},  090201.
%
\bibitem{Milosevic2006} 
Milošević, D. B.; Paulus, G. G.; Bauer, D.; Becker, W.
Above-threshold ionization by few-cycle pulses.
{\em J. Phys. B: At. Mol. Opt. Phys.} {\bf 2006}, {\em 39}, R203.
%
\bibitem{Arbo2010}
Arbó, D. G.; Ishikawa, K. L.; Schiessl, K.; Persson, E.; Burgdörfer, J.
Intracycle and intercycle interferences in above-threshold ionization: The time grating
{\em Phys. Rev. A} {\bf 2010}, {\em 81}, 021403(R).
%
\bibitem{Gramajo2018}
Gramajo, A. A.; Della~Picca, R.; López, S. D.; Arbó, D. G. 
Intra- and intercycle interference of angle-resolved electron emission in laser-assisted XUV atomic ionization.
{\em J. Phys. B: At. Mol. Opt. Phys.} {\bf 2018}, {\em 51},  055603.
%
\bibitem{DellaPicca2020}
Della~Picca, R.; Ciappina, M. F.; Lewenstein, M.; Arbó, D. G.
Laser-assisted photoionization: Streaking, sideband, and pulse-train cases.
{\em Phys. Rev. A} {\bf 2020}, {\em 102}, 043106.
%
\bibitem{Macri2003}
Macri, P. A.; Miraglia, J. E.; Gravielle, M. S.
Ionization of hydrogen targets by short laser pulses.
{\em J. Opt. Soc. Am. B} {\bf 2003} {\em 20}, 1801--1806.
%
\bibitem{DellaPicca2013},
Della~Picca, R.; Fiol, J.;Fainstein, P. D.
Factorization of laser–pulse ionization probabilities in the multiphotonic regime.
{\em J. Phys. B: At. Mol. Opt. Phys.} {\bf 2013}, {\em 46}, 175603.
%
\bibitem{symphonySFA}
Amini, K.; {et. al.}
Symphony on strong field approximation.
{\em Rep. Progr. Phys.} {\bf 2019}, {\em 82}, 116001.
%
\bibitem{Duchateau2002}
Duchateau, G.; Cormier, E.; Gayet, R. 
Coulomb-Volkov approach of ionization by extreme-ultraviolet laser pulses in the subfemtosecond regime.
{\em Phys. Rev. A} {\bf 2002}, {\em 66}, 023412.
%
\bibitem{Gramajo2016}
Gramajo, A. A.; Della~Picca, R.; Garibotti, C. R.; Arbó, D. G.
Intra- and intercycle interference of electron emissions in laser-assisted XUV atomic ionization.
{\em Phys. Rev. A} {\bf 2016}, {\em 94}, 053404.
%
\bibitem{Lopez2024}
López, S. D.; Ocello, M. L.; Arbó, D. G.
Time-dependent theory of reconstruction of attosecond harmonic beating by interference of multiphoton transitions.
{\em Phys. Rev. A} {\bf 2024}, {\em 110}, 013104. 
%
\bibitem{Ocello2025}
Ocello, M. L.; López, S. D.; Barlari, M.; Arbó, D. G. 
Time-Dependent Theory of Electron Emission Perpendicular to Laser Polarization for Reconstruction of Attosecond Harmonic Beating by Interference of Multiphoton Transitions. 
{\em Atoms} {\bf 2025}, {\em 13}, 99.
%
\end{thebibliography}
\end{document}